\documentclass[aps,prmaterials,reprint,superscriptaddress,floatfix,longbibliography]{revtex4-2}

\usepackage{amsmath,amssymb,graphicx,array,xcolor,soul,tikz,longtable}
\newcommand\tikznode[2]{\tikz[remember picture,baseline=(#1.base)]{\node(#1)[inner sep=-2pt]{#2};}}

\setcitestyle{super}

\def\e{\epsilon}
\newcommand{\mob}{cm$^2$/Vs}
\def\bk{{\bf k}}
\def\bq{{\bf q}}

\def\b{{\beta}}
\def\D{\partial}
\def\DE{\partial_{E_\b}}
\def\d{\delta}

\def\w{\omega}

\def\ve{\varepsilon}
\def\gmn{g_{mn\nu}(\bk,\bq)}
\def\gtwo{|g_{mn\nu}(\bk,\bq)|^2}
\def\sumq{\sum_{m\nu} \!\int\!\! \frac{d\bq}{\Omega_{\rm BZ}}}
\def\fone{f_{n\bk}\,}
\def\fonez{f^0_{n\bk}\,}
\def\ftwo{f_{m\bk+\bq}\,}
\def\ftwoz{f^0_{m\bk+\bq}\,}
\def\n{\,n_{\bq\nu}\,}
\def\deltap{\,\d(\ve_{n\bk} - \ve_{m\bk+\bq} + \hbar\w_{\bq\nu})\,}
\def\deltam{\,\d(\ve_{n\bk} - \ve_{m\bk+\bq} - \hbar\w_{\bq\nu})\,}
\def\time{\tau_{n\bk}}

\def\utoden{Oden Institute for Computational Engineering and Sciences, The University of Texas at Austin, Austin, Texas 78712, USA}
\def\utphysics{Department of Physics, The University of Texas at Austin, Austin, Texas 78712, USA}
\def\ucsb{Materials Department, University of California, Santa Barbara, California 93106-5050, USA}
\def\lanld{Theoretical Division, Los Alamos National Laboratory, Los Alamos, 87545 New Mexico, USA}
\def\lanlc{Center of Nonlinear Studies, Los Alamos National Laboratory, Los Alamos, 87545 New Mexico, USA}

\begin{document}

\title{Design of high-mobility \textit{p}-type GaN via the piezomobility tensor}


\author{Jie-Cheng Chen}
\affiliation{\utphysics}
\affiliation{\utoden}

\author{Joshua Leveillee}
\affiliation{\utoden}
\affiliation{\lanld}
\affiliation{\lanlc}

\author{Chris G. Van de Walle}
\affiliation{\ucsb} 

\author{Feliciano Giustino}
\email{fgiustino@oden.utexas.edu}
\affiliation{\utphysics}
\affiliation{\utoden}

\date{\today}

\begin{abstract}
Gallium nitride (GaN) is a wide-bandgap semiconductor of significant interest for applications in solid-state lighting, power electronics, and radio-frequency amplifiers. An important limitation of this semiconductor is its low intrinsic hole mobility, which hinders the development of \textit{p}-channel devices and the large-scale integration of GaN CMOS in next-generation electronics. Prior research has explored the use of strain to improve the hole mobility of GaN, but a systematic analysis of all possible strain conditions and their impact on the mobility is lacking. In this study, we introduce a piezomobility tensor notation to characterize the relationship between applied strain and hole mobility in GaN. To map the strain-dependence of the hole mobility, we solve the \textit{ab initio} Boltzmann transport equation, accounting for electron-phonon scattering and GW quasiparticle energy corrections. We show that there exist three optimal strain configurations, two uniaxial strains and one shear strain, that can lead to significant mobility enhancement. In particular, we predict room-temperature hole mobility of up to 164~\mob\ for 2\% uniaxial compression and 148~\mob\ for 2\% shear strain. Our methodology provides a general framework for investigating strain effects on the transport properties of semiconductors from first principles.
\end{abstract}

\maketitle

\section{Introduction}

Wurtzite gallium nitride (GaN) has played a fundamental role in the development of solid-state lighting, and is currently used in power electronics and wireless communications.\cite{Ponce1997,Flack2016,Amano2018,Ishida2013,Nakamura:1993,Nakamura:1995,Li:2016,Bader2018,Li:22} Owing to its wide bandgap of 3.4~eV, GaN offers lower leakage current and higher critical field as compared to conventional silicon, thus enabling reduced energy dissipation, faster switching, and higher operating temperatures.\cite{Amano2018,Flack2016}

GaN exhibits high electron mobility, in excess of 1000~\mob\ at room temperature for nominally-undoped single crystal samples,\cite{Nakamura:1992,Gotz:1996,Gotz:1998,Kyle:2014} which is comparable to the electron mobility in silicon.\cite{Flack2016} However, the room-temperature mobility of holes in GaN is only 30~\mob.\cite{Flack2016,Rubin:1994,Kozodoy:1998,Look1999,Cheong2000,Kozodoy:2000,Cheong2002,Arakawa:2016,Horita:2017,Hamaguchi} This imbalance hinders the development of complementary metal-oxide-semiconductors (CMOS) field-effect transistors and their large-scale integration in integrated circuits.

One potential avenue to improve the hole mobility of GaN is to reduce the hole effective masses via strain engineering.\cite{Suzuki:1996,Yeo:1998,Gupta:2019,Jena2019,Ponce:2019_2,Leveillee:2022} The rationale for this approach is twofold: (i) the dominant hole-phonon scattering mechanism in GaN is acoustic deformation potential (ADP) scattering,\cite{Ponce:2019_2} which is sensitive to the electronic density of states and hence the band effective masses; (ii) the topmost valence bands are heavy holes with an effective mass $m^*_{\rm hh}=1.94\,m_e$ along the $c$-axis direction \cite{Ponce:2019} ($m_e$ is the free electron mass). Therefore, reducing the hole effective mass would simultaneously increase the numerator and decrease the denominator in Drude's formula $\mu = e \tau / m^*$, where $\mu$, $e$, $\tau$ are the mobility, electron charge, and relaxation time, respectively.

Several theoretical studies investigated potential strategies to improve the hole mobility of GaN via strain engineering. In Refs.~\citenum{Ponce:2019_2,Ponce:2019}, using the \textit{ab initio} Boltzmann transport equation (\textit{ai}BTE), the authors found that biaxial tensile strain in the basal plane or uniaxial compressive strain along the $c$-axis leads to the split-off ($sh$) holes being lifted above the heavy-holes ($hh$), with a corresponding enhancement of the mobility by 230\% for 2\% strain.
In Ref.~\citenum{Bader:2019}, the authors employed a semiempirical ${\bf k}\!\cdot\!{\bf p}$ Hamiltonian to show that a uniaxial compression is the best approach for optimizing in-plane hole mobility in the GaN/AlN 2D hole gas. 
In Ref.~\citenum{Leveillee:2022}, using the \textit{ai}BTE, the authors predicted that lattice-matching GaN to ZnGeN$_2$ or MgSiN$_2$ substrates would induce an increase in the hole mobility by 50\% and 260\%, respectively. Additional efforts in this direction have been reported in Refs.~\citenum{Sun2007,Fu2009,Kuroiwa2019,Liu:2022,Miyazaki_2023,Li:2023,Cao:2023}. All these studies support the notion that strain engineering offers a promising avenue to increase the mobility of \textit{p}-type GaN. However, a broad and systematic investigation of possible strain configurations has not been performed yet.

\begin{figure}
\includegraphics[width=0.48\columnwidth]{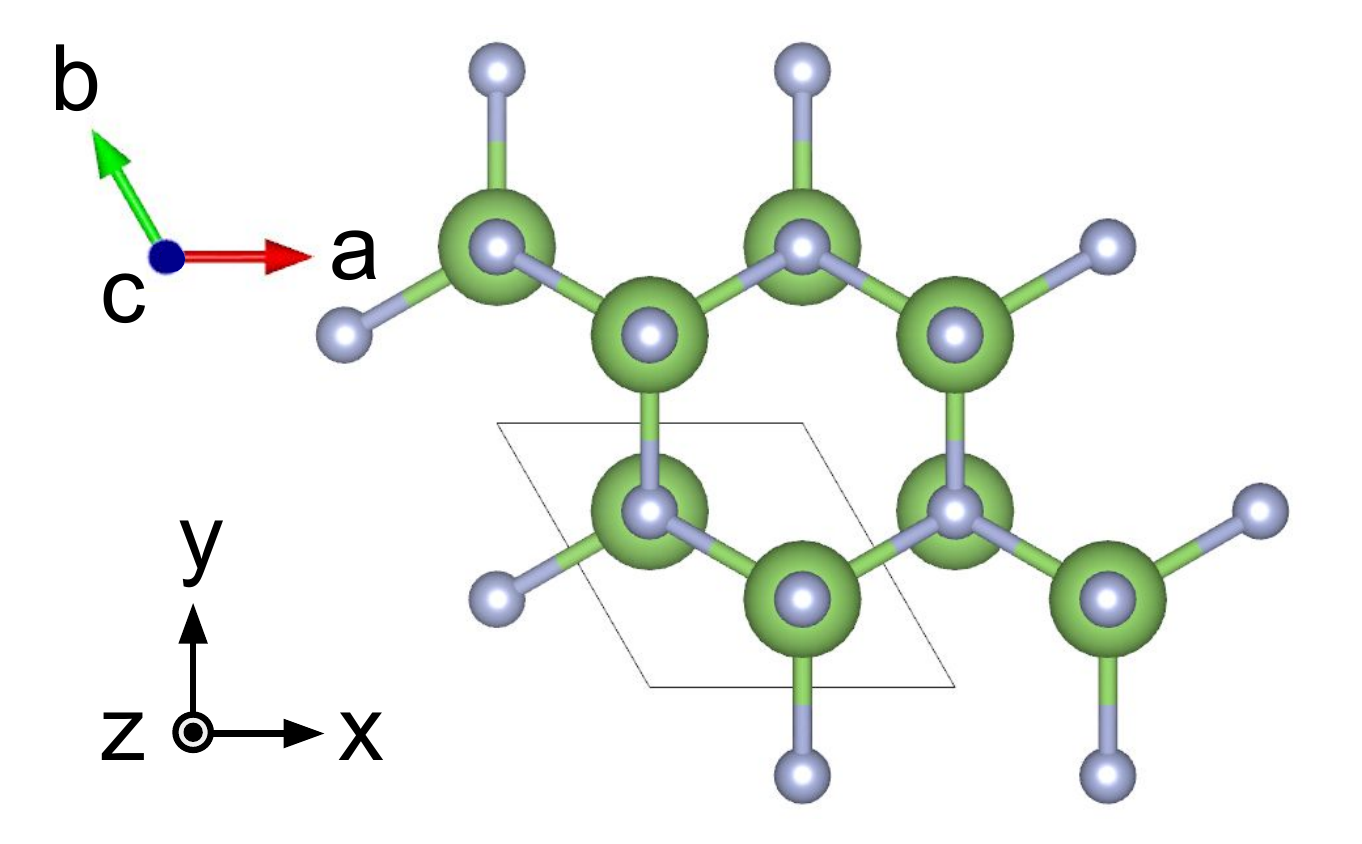}
\includegraphics[width=0.48\columnwidth]{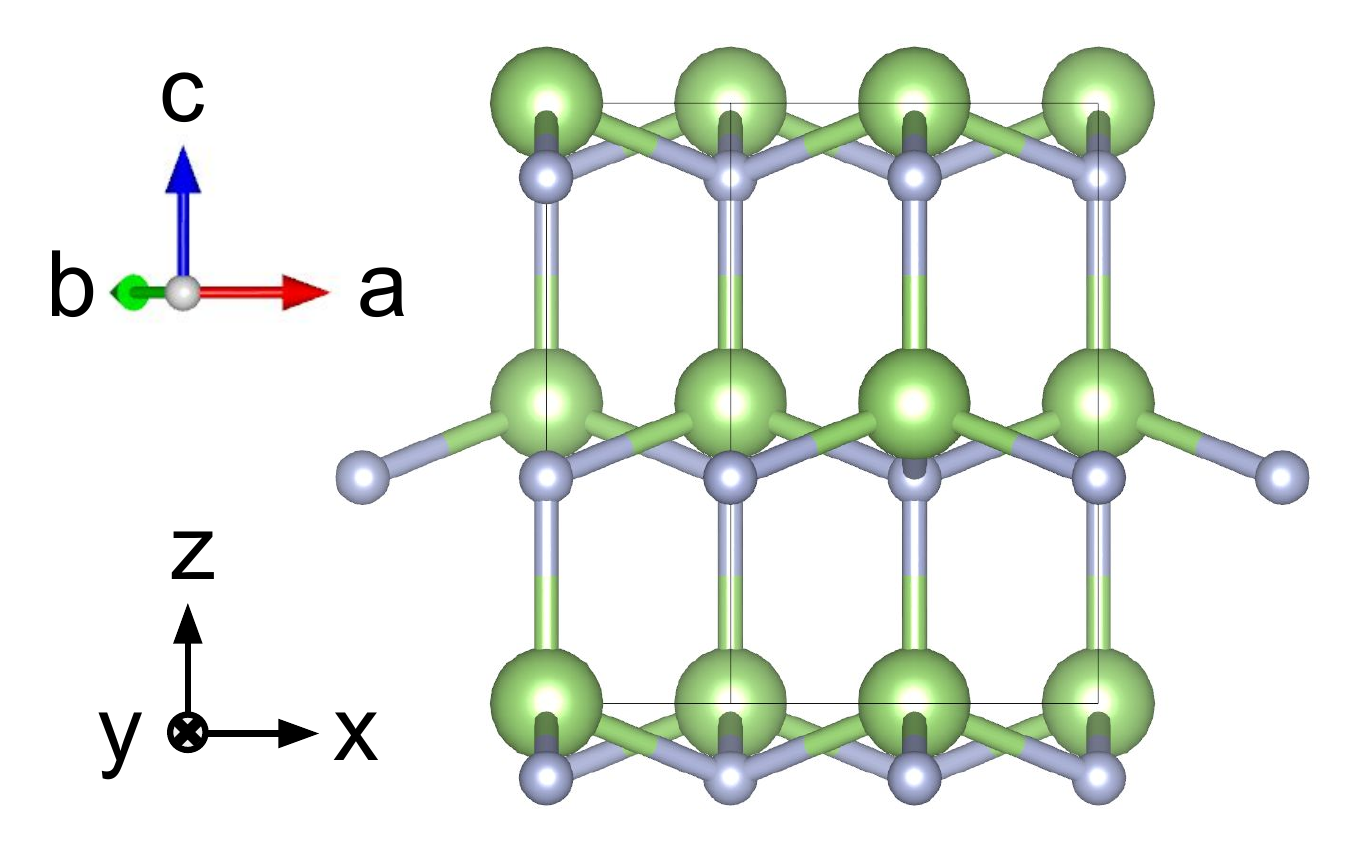}
\caption{Top view and side view of the crystal structure of wurtzite GaN, with Ga in green and N in gray. The primitive lattice vectors and the convention for the Cartesian reference frame employed throughout this manuscript are shown.}
\label{fig:struct}
\end{figure}

To fill this gap, we here introduce the piezomobility tensor as a conceptual tool for examining the effect of strain on carrier transport in a systematic manner. We evaluate this tensor by performing first-principles calculations of the phonon-limited carrier mobility in GaN via the \textit{ai}BTE, and we include spin-orbit coupling (SOC) and GW quasiparticle corrections for high-accuracy results. We find three optimal strain configurations, two corresponding to uniaxial strain and one to shear strain, that induce a significant enhancement of the theoretical hole mobility. Specifically, we obtain a room-temperature hole mobility of up to 164~\mob\ for a configuration corresponding to single-component 2\% compressive $\e_3$ strain (uniaxial, $c$-axis; in Voigt notation, see Fig.~\ref{fig:struct} for crystal orientation). In specifying this uniaxial compression, we keep all other strain components equal to zero. Experimentally, such single-component strain is not achievable, but as discussed in Sec.~\ref{ssec.multi} we find that it constitutes a very good approximation to what can be obtained in realistic conditions. The second optimal configuration corresponds to $\e_1$ (uniaxial, $x$-axis), resulting in 159~\mob; and the third to $\e_4$ ($yz$ shear) strain, resulting in 148~\mob. The first of these strain configurations was already discussed in the literature,\cite{Ponce:2019_2,Ponce:2019} while the other two configurations are new. To the best of our knowledge, the use of shear strain has not been discussed before. We also rationalize our findings in terms of the band effective mass, and find that the conductivity effective mass constitutes a reliable approximate descriptor of the change of hole mobility under strain, both single- and multi-component.

The manuscript is organized as follows. In Sec.~\ref{sec.pz} we introduce the notion of piezomobility tensor and state the properties of this tensor for wurtzite GaN. Section~\ref{sec.comp} summarizes the computational setup employed in this work, as well as the key components of the \textit{ai}BTE methodology. In Sec.~\ref{sec.results} we describe our results. In particular, Sec.~\ref{ssec.summary} provides a concise overview of our findings for single-component strain, and Sec.~\ref{ssec.indepth} analyzes the best case scenarios identified here in greater detail. In Sec.~\ref{ssec.nonlin} we investigate the importance of effects beyond the linear order in the strain, and in Sec.~\ref{ssec.mass} we show how the hole mobility enhancement in GaN primarily arises from the reduction of the hole effective mass. Section~\ref{ssec.multi} considers a few realistic strain scenarios where multiple strain components are simultaneously nonvanishing. In Sec.~\ref{sec.concl}, we discuss our results in the light of recent experimental work and we draw our conclusions.

\section{Piezomobility tensor}\label{sec.pz}

To map out the relation between strain and mobility, we make the observation that both the strain tensor and the mobility tensors are symmetric. In fact, the strain tensor $\e_{\alpha\beta}$ with $\alpha,\beta$ being Cartesian directions is symmetric by definition.\cite{Nye:1985} The diagonal elements of this tensor describe axial strains, while the off-diagonal components characterize shear strains. Furthermore, the mobility tensor $\mu_{\alpha\beta}$ is also a symmetric tensor, as can be shown using Onsager's reciprocity relations.\cite{Garrod:1983} The symmetry of the mobility tensor stems from the fundamental time-reversal invariance of the microscopic dynamics. As both $\e_{\alpha\beta}$ and $\mu_{\alpha\beta}$ are symmetric, they each possess 6 independent components, which can conveniently be indexed via Voigt's notation: $xx \leftrightarrow 1$, $yy \leftrightarrow 2$, $zz \leftrightarrow 3$, $yz \leftrightarrow 4$, $xz \leftrightarrow 5$, $xy \leftrightarrow 6$. With this notation, our rank-2 tensors are rearranged into vectors $\e_i$ and $\mu_i$, with $i=1,2,\cdots 6$. The most general relation between the components $\e_i$ and $\mu_i$ for small strain is given by:
  \begin{equation}\label{eq.K1}
    \Delta\mu_{i}=K_{ij}\e_{j},
  \end{equation}
where $\Delta \mu_i$ refers to the change with respect to the ground-state unstrained crystal, and summation over repeated indices is implied. We will refer to the coefficient matrix $K$ as the \textit{piezomobility tensor}. The element $K_{ij}$ of this tensor corresponds to the change of the $i$-th component of the mobility tensor under the effect of the $j$-th component of the strain tensor. Crystal symmetry dictates the allowed components of the piezomobility tensor; in the case of wurtzite GaN, with space group P6$_3$mc (Fig.~\ref{fig:struct}), this tensor has 20 independent components, two components are identical, and 15 components vanish. The relation between the mobility and strain components using Cartesian directions, as well as a graphical representation of the piezomobility tensor, are as follows: 
  \begin{equation}\label{eq.K2}
    \begingroup
    \renewcommand*{\arraystretch}{0.95}
    \begin{pmatrix}
      \Delta \mu_{xx} \\ \Delta \mu_{yy} \\ \Delta \mu_{zz} \\
      \Delta \mu_{yz} \\ \Delta \mu_{xz} \\ \Delta \mu_{xy} 
    \end{pmatrix} = K 
    \begin{pmatrix}
      \epsilon_{xx} \\ \epsilon_{yy} \\ \epsilon_{zz} \\
      \epsilon_{yz} \\ \epsilon_{xz} \\ \epsilon_{xy} 
    \end{pmatrix},
    \quad
    K=\begin{pmatrix}
      \bullet & \bullet & \tikznode{a}{$\bullet$} & \bullet & \bullet & \bullet \\
      \bullet & \bullet & \tikznode{b}{$\bullet$} & \bullet & \bullet & \bullet \\
      \bullet & \bullet & \bullet & \bullet & \bullet & \bullet \\
      &  &  & \bullet &  &  \\
      &  &  &  & \bullet &  \\
      &  &  &  &  & \bullet 
    \end{pmatrix}.
    \endgroup
  \end{equation}
  \tikz[overlay,remember picture,line width=1pt]{\draw[-](a)--(b)}%
Here, the $z$ direction is aligned with the $c$-axis, dots represent allowed tensor components, and the connecting line indicates that two components are identical by symmetry. 

The matrix $K$ in Eq.~\eqref{eq.K2} is a close relative of the piezoresistance tensor $\pi_{ij}$.\cite{Barlian:2009,Smith:1954} This latter tensor is defined as the relative change of resistivity $\rho_i$ in response to the stress $\sigma_j$ (both in Voigt's notation), and therefore it incorporates information about the carrier density: $\Delta \rho_i/\rho_i =\pi_{ij}\sigma_j$. By contrast, the piezomobility tensor of Eq.~\eqref{eq.K1} is defined in terms of \textit{strain} and carrier \textit{mobility}. This definition is advantageous in the context of materials design, since (i) unlike the conductivity, the carrier mobility is an intrinsic property of the undoped semiconductor, and (ii) in \textit{ab initio} calculations, strain can directly be imposed by simply specifying the lattice vectors.
We emphasize that the linear relation in Eq.~(1) is only valid at very
small strain. In Sec.~IV we discuss how to address the more general cases where nonlinear terms become important, using the conductivity effective mass.

\section{Computational methods}\label{sec.comp}

\subsection{Calculation setup}

All calculations are performed within the framework of density functional theory (DFT) and GW perturbation theory using the Quantum ESPRESSO package,\cite{Giannozzi:2017} the EPW code,\cite{FG:2007,Lee:2023}, the Wannier90 code,\cite{Mostofi2014} and the BerkeleyGW code.\cite{BGW}

We employ the local density approximation to DFT exchange and correlation (DFT-LDA),\cite{LDA} and we include spin-orbit coupling throughout via fully-relativistic optimized norm-conserving Vanderbilt pseudopotentials.\cite{ONCV} For higher accuracy, we include Ga semicore states in all calculations. A plane-wave kinetic energy cutoff of 120\,Ry is used throughout for Kohn-Sham wavefunctions.

Starting from the optimized ground-state structure of wurtzite GaN (stress below 0.02\,kbar, Table~\ref{tab:alat}), we construct strained lattices and optimize the atomic coordinates at fixed lattice vectors (until individual forces fall below 10$^{-4}$\,Ry/Bohr). In ground state calculations, we sample the Brillouin zone using an 8$\times$8$\times$8 $\Gamma$-centered $\bk$-mesh. We compute phonon dispersions for every strain state using density functional perturbation theory (DFPT).\cite{Baroni:2001} For improved interpolation quality of electronic band structures and electron-phonon matrix elements, we calculate phonons on slightly denser 10$\times$10$\times$10 $\bq$-grids. We employ dipole corrections for the matrix elements~\cite{Verdi:2015}, but we do not include quadrupole corrections~\cite{Brunin:2020,Jhalani:2020} which have been shown to correct mobilities upwards by 13\% in unstrained GaN.~\cite{Leveillee:2022}

One-shot GW calculations are performed to account for many-body correlation effect beyond DFT.\cite{HybertsenLouie} We use 800 spinor states to converge GW quasiparticle corrections, and a kinetic energy cutoff of 15\,Ry for the dielectric matrices. Quasiparticle eigenvalues are computed on 10$\times$10$\times$10 $\bk$-grids.

The intrinsic phonon-limited hole mobility is computed using the linearized Boltzmann transport equation,\cite{Ponce:2018,Ponce2021} which we refer to as \textit{ai}BTE.  Ultrafine sampling of the Brillouin zone for Boltzmann transport calculations is achieved via Wannier-Fourier interpolation of electron bands, phonon dispersions, and electron-phonon coupling matrix elements, as implemented in the EPW code,\cite{FG:2007,Lee:2023} which calls the Wannier90 code in library mode.\cite{Mostofi2014} Wannier functions are obtained starting from N-$2p$ initial projections for the 12 spinor valence bands. Convergence of the room-temperature hole mobility is reached using 80$\times$80$\times$80 $\bk$- and $\bq$-point grids, using adaptive smearing.\cite{Ponce2021} To accelerate calculations, only electronic states within 200\,meV from the valence-band maximum are considered. In all strain configurations considered in this work, the Fermi level lies slightly above the valence band top, therefore this energy window yields converged results in all cases. 
The conductivity effective mass tensors [Eq.~\eqref{eq.cem} below] are computed on the same 80$\times$80$\times$80 fine $\bk$-grids employed for hole mobility calculations.

To consider the multi-component strains in GaN due to spontaneous strain relaxation, we calculate the elastic constants of GaN. Elastic constants are defined as the proportionality coefficients between stress and strain,
\begin{equation}
    \sigma_i = \sum_{j} C_{ij} \epsilon_j,
\end{equation}
where the stress $\sigma_i$ and strain $\e_j$ are expressed in Voigt's notation. Since stress vanishes in the unstrained structure, the elastic constants are obtained from the above equation by imposing the single-component strain $\e_j$ to the unit cell, relaxing the atomic coordinates, and filling the $j$-th column of $C_{ij}$ with the calculated stress $\sigma_i$ in this configuration. The stress is calculated via the stress theorem,\cite{Nielsen1985} as implemented in Quantum ESPRESSO.\cite{Giannozzi:2017}

\begin{table}
\caption{Lattice parameters of wurtzite GaN: Comparison between this work and literature data.}
\label{tab:alat}
\begin{tabular}{ p{4.5cm} m{1.8cm} m{1.8cm}}
\hline\hline\\[-7pt]
Lattice Parameters (\AA) & $a$ & $c$ \\
\hline \\[-6pt]
This work & 3.155 &  5.141  \\
Expt., Ref.~\citenum{Tellekamp} & 3.186 &  5.160 \\
Expt., Ref.~\citenum{Qian1996} & 3.189 &  5.178   \\
Theo., Ref.~\citenum{Leveillee:2022} & 3.146 &  5.136 \\[2pt]
\hline\hline
\end{tabular}
\end{table}

\subsection{Boltzmann transport equation and conductivity effective mass}

To keep this manuscript self-contained, we briefly summarize the key equations used in \textit{ai}BTE calculations.
The carrier mobility tensor is evaluated as:\cite{Ponce:2020}
  \begin{eqnarray}
   && \mu_{\alpha\beta} = -\frac{e}{\Omega} \sum_n\int \frac{d\bk}{\Omega_{\rm BZ}} v_{n\bk,\alpha} \DE \fone~,
  \end{eqnarray}
where $e$ is the electron charge, $\Omega$ is the volume of the unit cell, $v_{n\bk,\alpha}$ is the band velocity in the Cartesian direction $\alpha$, and $\DE \fone= (\D \fone/\D E_\b)_{E_\b=0}$ is the linear response of the carrier occupation $f_{n\bk}$ to the component of the applied electric field along the Cartesian direction $\beta$. To calculate the intrinsic phonon-limited hole mobility, we compute $\DE \fone$ by iteratively solving:
  \begin{eqnarray}
   \DE \fone &=& 
  e \left. \frac{\D f^0}{\D\ve}\right|_{\ve_{n\bk}} \!\!\!\!v_{n \bk,\b} \time \nonumber \\
  &
  +&\frac{2\pi\time}{\hbar} \sumq  \gtwo \nonumber \\
  & \times &[
  (1-\fonez+\n) \deltap  \nonumber \\ &+& (\fonez + \n) \deltam ] \nonumber \\ &\times& \DE\ftwo, 
  \end{eqnarray}
where $\gmn$ are the electron-phonon matrix elements, $\fonez$ is the Fermi-Dirac electron occupation, $\n$ is the Bose-Einstein phonon occupation, $\ve_{n\bk}$ and $\w_{\bq\nu}$ are the electron energy and phonon frequency, respectively. The total scattering rate $1/\time$ of a Bloch state with band index $n$ and wave vector $\bk$ is given by:
  \begin{eqnarray}
   \frac{1}{\time} &=& \frac{2\pi}{\hbar} 
  \sumq \gtwo\nonumber \\
  & \times &[ (1-\ftwoz+\n)\deltam \nonumber \\ &+& (\ftwoz+\n)\deltap], 
  \end{eqnarray}
and corresponds to the imaginary part of the Fan-Migdal self-energy (after multiplying by $\hbar/2$).\cite{FG:RMP}

The effective mass tensor $m^*_{n\bk,\alpha\beta}$ of band $n$ at the wavevector $\bk$ is defined by the relation:
 \begin{equation}
  (m^*_{n\bk})^{-1}_{\alpha\beta} = \frac{1}{\hbar^2}\frac{\D^2\ve_{n\bk}}{\D k_{\alpha}\D k_{\beta}}.
  \end{equation}
In order to take into account the carrier distribution in mobility calculations, as determined by the electronic density of states and the Fermi-Dirac occupations, we evaluate the following average of the above effective mass, which was referred to as a ``conductivity effective mass'' in Ref.~\citenum{Hautier2014}:
  \begin{equation}
  (m^*_{\rm c})^{-1}_{\alpha\beta} = \frac{\Omega_{\rm BZ}^{-1}\,\sum_{n} \!\int d\bk \,(m^*_{n\bk})^{-1}_{\alpha\beta}
     \,f_{n\bk}}{\Omega_{\rm BZ}^{-1}\,\sum_{n} \int d\bk \, f_{n\bk}}.
  \end{equation}
From this definition, it is immediate to verify that the conductivity effective mass relates to the carrier mobility in the constant relaxation time approximation (CRTA) by:
  \begin{equation}\label{eq.cem}
  (m^*_{\rm c})_{\alpha\beta} = e \tau (\mu_{\rm CRTA}^{-1})_{\alpha\beta},
  \end{equation}
where $\tau$ is the constant relaxation time. In this work, we evaluate the conductivity effective mass using Eq.~\eqref{eq.cem}.

\section{Results}\label{sec.results}

\subsection{Summary of key findings}\label{ssec.summary}

\begin{figure}
\includegraphics[width=0.95\columnwidth]{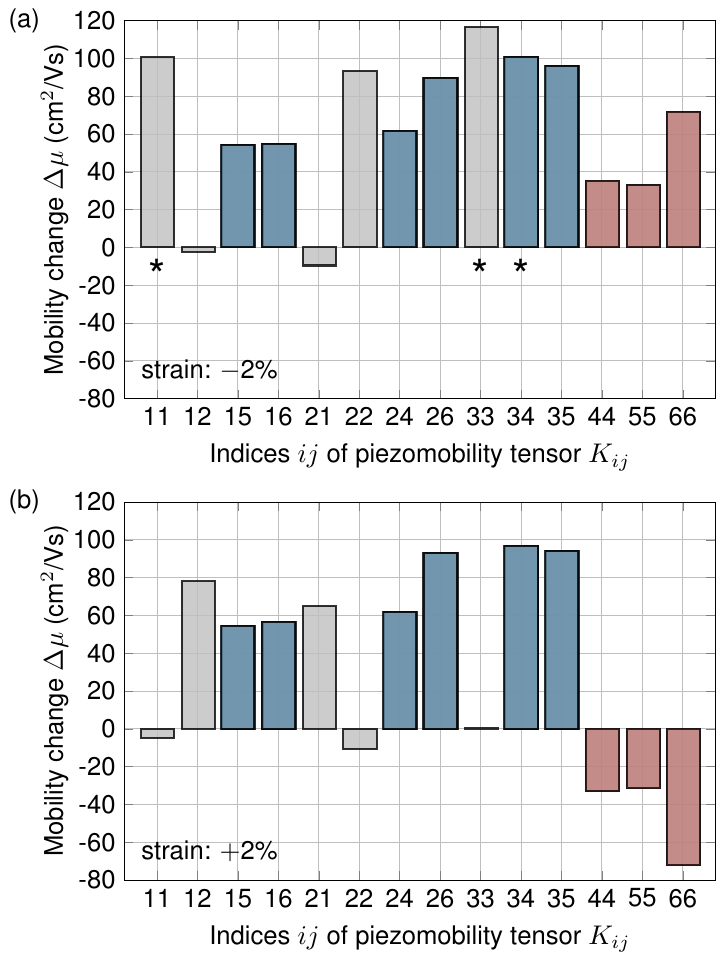}
\caption{(a) Change of the room-temperature hole mobility of GaN under 2\% compressive single-component strain with all other strain components equal to zero, for each component of the piezomobility tensor $K_{ij}$. For example, the entry ``12'' indicates the value of $\Delta \mu_1$ resulting from the strain $\epsilon_2 = -0.02$, in Voigt notation. Entries for which the mobility change is smaller than 15\,\mob\ for both tensile and compressive strain are omitted for clarity. Gray bars indicate uniaxial strain; blue and red bars are for the diagonal and off-diagonal components of the mobility under shear strain, respectively. \textit{ai}BTE calculations performed using GW quasiparticle band structures, for a carrier concentration of 10$^{18}$\,cm$^{-3}$. (b) Same as (a), but for 2\% tensile strain. Asterisks denote the single-component strains discussed in the text.}
\label{bar:strn2}
\end{figure}

In Fig.~\ref{bar:strn2} we provide a summary view of the piezomobility tensor calculated for 2\% strain, in the form of a bar chart. By ``2\% strain'' we mean a single-component strain, with all other strain components set to zero. The \textit{ai}BTE calculations are performed for the hole mobility of intrinsic GaN at room temperature. Data in the figure are obtained using GW quasiparticle band structures, and are reported for completeness in Table~\ref{tableIII} in Appendix alongside with the corresponding DFT-LDA values. Following the definition of the piezomobility tensor in Eq.~\eqref{eq.K1}, in this figure we consider single-component strain, meaning that we set up a strained crystal unit cell where the strain tensor has only one nonzero component (or two in the case of off-diagonal components). In realistic situations, multiple strain components are usually nonzero due to Poisson's effect. We defer a discussion of multi-component strain to Sec.~\ref{ssec.multi}. We initially focus on 2\% tensile and compressive strain solely for illustration purposes. This choice is motivated by the observation that achieving strain levels up to 3\% is currently feasible via epitaxial growth.\cite{Jain2000,Wagner2002,Li2014,Qi2015,Qi2017,Jena2019} Additional strain levels are examined in Sec.~\ref{ssec.nonlin}.

Figure~\ref{bar:strn2} shows that, upon compression, single-component uniaxial strains ($K_{11}$, $K_{22}$, $K_{33}$) generate significant mobility enhancements in the same directions as the applied strain, up to a factor of $3\times$ (gray bars). Furthermore, the imposition of single-component shear strain enhances the diagonal components of the mobility along one of the two strain directions ($K_{15}$, $K_{16}$, $K_{24}$, $K_{26}$, $K_{34}$, $K_{35}$), with up to 3$\times$ mobility enhancement for both compressive and tensile strain (blue bars). The effect of shear strain on the off-diagonal components of the mobility tensor ($K_{44}$, $K_{55}$, $K_{66}$) is also relatively pronounced, although we find mobility enhancement for compressive strain and suppression for tensile strain (brown bars). 

\subsection{In-depth analysis of optimum strain configurations}\label{ssec.indepth}

\begin{figure}
\includegraphics[width=\columnwidth]{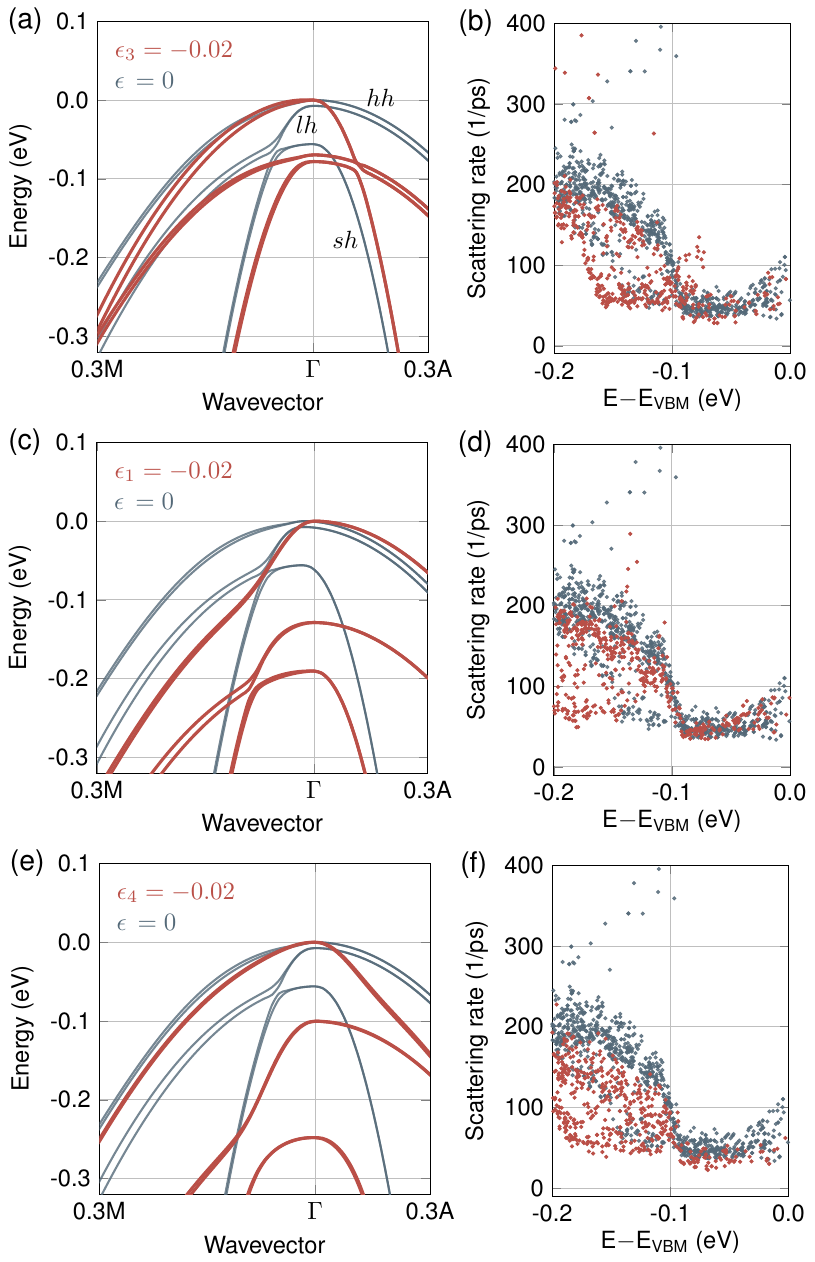}
\caption{(a) GW quasiparticle band structures of unstrained GaN (dark blue) and GaN under 2\% compressive uniaxial single-component strain along the $c$ axis ($\epsilon_3 = -0.02$, brown). The heavy hole, light hole, and split-off hole bands of the unstrained crystal are marked. (b) Hole scattering rates for the same systems shown in (a), using the same color code. The step-function-like feature near $-$0.1\,eV corresponds to the threshold for the emission of longitudinal optical phonons at 93\,meV. (c),(d) Same as for (a),(b) but for 2\% compressive uniaxial single-component strain along $x$ ($\e_1 = -0.02$). (e),(f) Same as for (a),(b) but for 2\% single-component $yz$ shear strain ($\e_4 = -0.02$). The band structure and scattering rates of GaN under each optimal strain configuration are shown in brown, while those of unstrained GaN are shown in dark blue for comparison.}
\label{fig:band_ph_tau}
\end{figure}

In the following, we analyze the three best-case strain configurations, which correspond to the single-component compressive strains $K_{33}$, $K_{11}$, and $K_{34}$. These cases are marked by asterisks in Fig.~\ref{bar:strn2}.

The coefficient $K_{33}$ corresponds to the highest mobility enhancement under compressive strain. It corresponds to the mobility response along the $c$ axis to single-component uniaxial compression along the same axis. For 2\% compression we calculate a mobility of 164\,\mob\ at room temperature. This result is consistent with a prior theoretical work, which attributed the effect to a reversal of crystal field splitting lifting the split-off holes above the heavy holes and light holes.\cite{Ponce:2019_2,Ponce:2019,Miyazaki_2023} 
In Fig.~\ref{fig:band_ph_tau}(a) we confirm this effect by showing how the split-off holes along the $\Gamma A$ high-symmetry line raise above the light and heavy hole bands by 75\,meV. As a result, the conductivity effective mass decreases by a factor $3\times$ from 1.65\,$m_e$ in the unstrained system to 0.52\,$m_e$ under 2\% strain ($m_e$ is the free electron mass, see Table~\ref{tableIV} in Appendix for all our calculated values). The reduced effective mass improves the mobility via two mechanisms: (i) the increase of the band velocity, and (ii) the reduction of the hole density of states and hence the scattering phase space. This latter effect leads to a reduction of the scattering rates, as is shown in Fig.~\ref{fig:band_ph_tau}(b). We also note that strain has only a minor effect on the phonon dispersions, see Fig.~\ref{fig:phSI} in Appendix.

The second strongest mobility enhancement is found for the $K_{11}$ coefficient, corresponding to single-component uniaxial compression along $x$ (Fig.~\ref{fig:struct}). For 2\% strain we obtain a room-temperature mobility $\mu_{xx}=159$\,\mob\ in the strain direction. Figure~\ref{fig:band_ph_tau}(c) shows that, in this case, the heavy holes and the split-off holes are pushed well below the light holes, leading to a reduction in the effective mass in the in-plane direction [$\Gamma M$ line in Fig.~\ref{fig:band_ph_tau}(c)]. As a result, also the conductivity effective mass decreases by a factor 3$\times$, from 1.32\,$m_e$ in the unstrained system to 0.41\,$m_e$ under strain. In this case, scattering rates for carriers with energy near room temperature are not significantly reduced, see Fig.~\ref{fig:band_ph_tau}(d). We point out that this same strain configuration was investigated in Ref.~\citenum{Bader:2019} using a ${\bf k}\cdot{\bf p}$ model for the 2D hole gas in a GaN/AlN heterostructure. These authors reported a mobility enhancement under strain in qualitative agreement with our results, although the numerical value of the mobility was probably underestimated due to the approximations employed in the model.

To our knowledge, the effect of shear strain on the mobility of GaN has not been reported hitherto. As shown in Fig.~\ref{bar:strn2}(a), the coefficient $K_{34}$ also corresponds to a significant mobility enhancement; this coefficient describes the response of $\mu_{zz}$ to the shear strain $\mu_{yz}$. The sign of the strain is inconsequential as the positive and negative strain configurations are related by an inversion of the $b$ axis and a translation by half the unit cell size along the $c$ axis [Fig.~\ref{bar:strn2}(b)]. In this scenario, we calculate a mobility $\mu_{zz}=146$\,\mob\ at room temperature, which is also a factor $3\times$ that of the unstrained crystal. Figure~\ref{fig:band_ph_tau}(e) shows that, in this case, the light holes and split-off holes are pushed down, leaving the heavy holes as the top of the valence manifold; at the same time, the heavy hole band is heavily warped along the $\Gamma A$ line, leading to a factor of 2$\times$ decrease in conductivity effective mass, from 1.65\,$m_e$ to {0.75\,$m_e$. As for the case of $\epsilon_1$ strain, the origin of higher mobility lies in the synergistic increase of higher band velocity and lower scattering rates [Fig.~\ref{fig:band_ph_tau}(f)] resulting from the reduced conductivity mass. For completeness, all our calculated mobility values are reported in Table~\ref{tableIII} in Appendix. Furthermore, in Sec.~IV.E we discuss how the mobility changes when going from these idealized single-component strain configurations to more realistic multi-component strain scenarios.

\subsection{Nonlinear effects}\label{ssec.nonlin}

\begin{figure}[b!]
\includegraphics[width=\columnwidth]{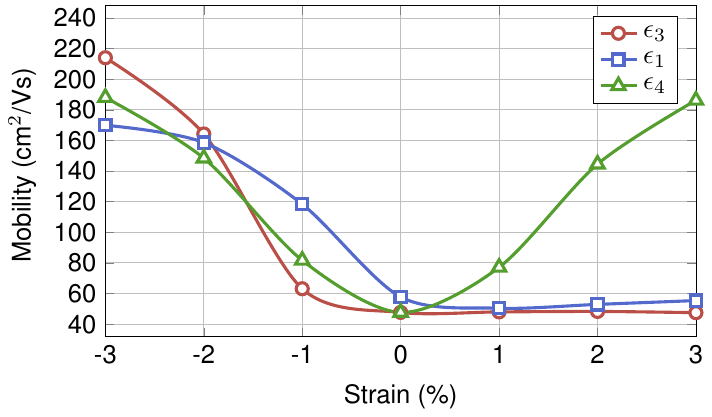}
\caption{Calculated room-temperature hole mobility of GaN as a function of strain. We report data for single-component $\e_3$ strain (uniaxial strain along $z$, red), single-component $\e_1$ strain (uniaxial strain along $x$, blue), and single-component $\e_4$ strain ($yz$ shear strain, green). In all cases we indicate the highest component of the mobility tensor ($\mu_{zz}$, $\mu_{xx}$, and $\mu_{zz}$, respectively). Lines are guides to the eye. \textit{ai}BTE calculations performed using GW quasiparticle band structures, for a carrier concentration of 10$^{18}$\,cm$^{-3}$ at room temperature. Note that the mobilities shown correspond to the best-case \textit{single-component} strain configurations. The effects of spontaneous strain relaxation are discussed in detail in Sec. IV.E.} 
\label{fig:best_case}
\end{figure}

So far we have presented results for 2\% strain levels. Since strains up to 3\% have been realized in experiments,\cite{Jain2000,Wagner2002,Li2014,Qi2015,Qi2017,Jena2019} in Fig.~\ref{fig:best_case} we examine the hole mobility of GaN for the single-component strains $\epsilon_3$, $\epsilon_1$, and $\epsilon_4$ discussed in Sec.~\ref{ssec.indepth}, this time by varying the strain from $-$3\% to $+$3\%. From this figure we see that, at 3\% strain, we find impressive room-temperature phonon-limited mobilities of up to 214\,\mob, suggesting that careful strain engineering holds the key for realizing high-mobility \textit{p}-type GaN.

Figure~\ref{fig:best_case} also shows that, for the strain levels considered in this work, the mobility vs.\ strain curves are highly nonlinear. This observation suggests that one should go beyond the linear piezomobility tensor $K_{ij}$ to describe these effects. However, higher-order expansions would introduce a large number of nonlinear piezomobility coefficients (e.g. 216 coefficients in the second-order expansion), making this analysis impractical. Since we are ultimately calculating mobilities under finite strains (rather than their derivatives with respect to the strains), it is advantageous to maintain the piezomobility notation in Eq.~\eqref{eq.K1} as an effective tool to systematically map out the strain-mobility relation. To estimate mobilities under highly-nonlinear strain conditions, we can employ the conductivity effective mass, as discussed in the next section.

\subsection{Role of conductivity effective mass}\label{ssec.mass}

In order to rationalize the trends observed in Fig.~\ref{fig:best_case}, we proceed to check whether also in this highly nonlinear regime the change in mobility is dominated by the variation of the conductivity effective mass. To this end, in Fig.~\ref{fig:effm} we examine the correlation between the diagonal components of the conductivity effective mass tensor and the corresponding components of the mobility (we do not consider off-diagonal components of the mobility as these vanish in the unstrained crystal). We obtain a Spearman's rank $\rho = -0.9$, indicating strong inverse correlation between mobility and mass. This observation is in line with the elementary Drude model, whereby $\mu = e \tau / m^*$, where $e$ is the electron charge and $\tau$ the characteristic relaxation time. A linear regression in log scale indicates that the best fit is obtained via 
\begin{equation}
\mu = 93/(m^*\!/m_e)^{0.89}\,\mbox{\mob},
\end{equation}
as shown by the brown line in Fig.~\ref{fig:effm}. The deviation of this formula from the Drude model can be ascribed to the effect of density of states mentioned earlier, which also correlates with the effective mass. For comparison, in Fig.~\ref{fig:effm} we also show that all data are bounded by a simple Drude formula with $\mu = (110\pm40)/(m^*\!/m_e)$\,\mob. These results suggest that (i) the nonlinearity observed in Fig.~\ref{fig:best_case} reflects the nonlinearity in the effective mass, and (ii) the conductivity effective mass is a key material descriptor to predict the effect of strain on the transport properties of \textit{p}-type GaN.

Besides the conductivity effective mass, the other key descriptor of the effect of strain on the mobility is the carrier scattering rate. The variation of the scattering rates with strain is precisely what leads to the spread of data points around the average line in Fig.~5. While this effect is non-negligible, the calculation of scattering rates requires the solution of the \textit{ai}BTE, which is significantly more demanding than the evaluation of the effective mass. Therefore, we propose that the effective mass should be the first point of call for a preliminary assessment of the strain-induced change of mobility.

The strong correlation between effective mass and mobility also provides us with a key to rationalize the difference between GW and DFT-LDA calculations of the hole mobility. All data presented thus far employ GW quasiparticle band structures. For completeness, in Fig.~\ref{fig:gw_dft}(a) we show a comparison between mobility values obtained from GW or DFT-LDA bands. It is seen that GW calculations yield consistently higher mobilities, with an average enhancement of 36\% over DFT-LDA calculations. In Fig.~\ref{fig:gw_dft}~(b) we compare the corresponding conductivity effective masses; here we see that GW yields masses that are 90\% lighter than in DFT-LDA on average. Since hole scattering in GaN is dominated by acoustic phonons,\cite{Ponce:2019_2} the mobility should scale with the mass as $\mu \!\propto\! (m^*)^{-5/2}$; by taking $(90\%)^{-5/2}$ we would expect a mobility enhancement of 30\%, which is close to the value reported above.
 These comparisons indicate that GW calculations tend to yield higher mobility than DFT-LDA as a result of the lighter effective masses. Furthermore, these findings reinforce the notion that the conductivity effective mass is the key descriptor for engineering high-mobility \textit{p}-type GaN, irrespective of the computational method employed. For completeness, all our calculated GW conductivity effective masses are reported in Table~\ref{tableIV} in Appendix.

\begin{figure}
\includegraphics[width=\columnwidth] {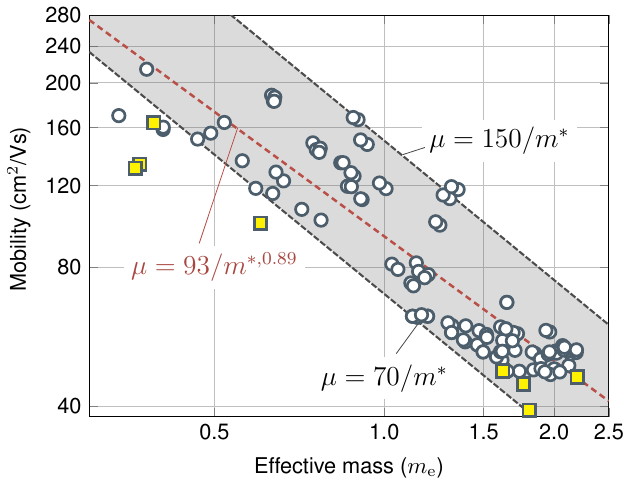}
\caption{Correlation between hole mobility of strained GaN and conductivity effective mass. Each data point in white circles corresponds to a pair $(m^*_{\alpha\alpha},\mu_{\alpha\alpha}), \alpha=1,\cdots 3$, as calculated for the strain configurations considered in Fig.~\ref{fig:best_case}. The brown line indicates the function $\mu = 93/m^{*,0.89}$\,\mob\ in log scale, with $m^*$ in units of $m_e$, as obtained from linear regression. The dashed black lines indicate the curves $\mu^* = 70/m^*$\,\mob\ and $\mu^* = 150/m^*$\,\mob, and the shading is a guide to the eye. Data points in yellow squares correspond to the four multi-component strain configurations analyzed in Sec.~IV.E.}
\label{fig:effm}
\end{figure}

\begin{figure}
\includegraphics[width=\columnwidth] {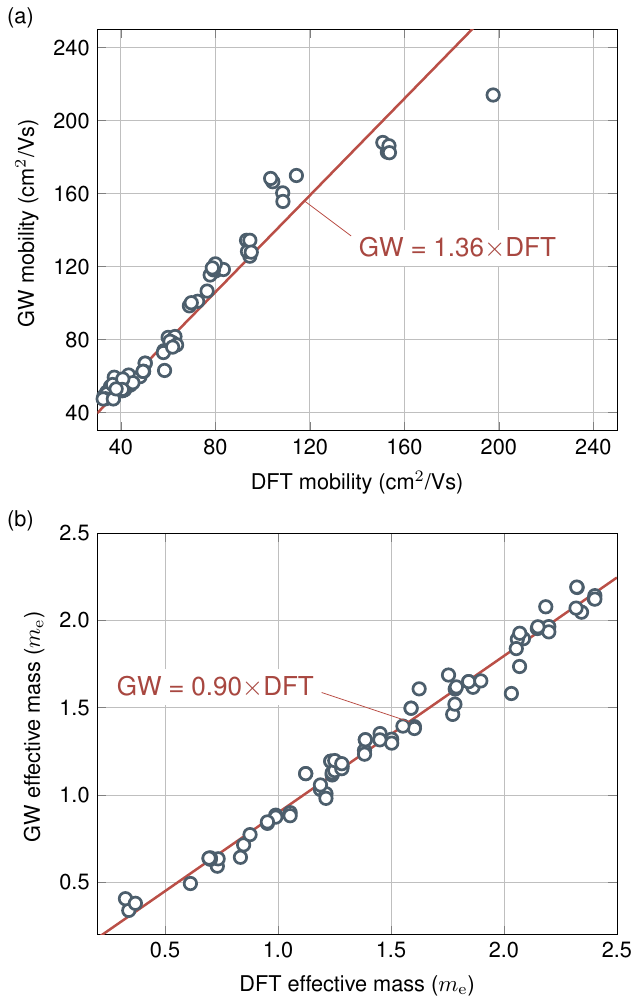}
\caption{Component-wise comparison of (a) GW and DFT-LDA hole mobilities, and (b) GW vs.\ DFT-LDA conductivity effective mass. Each data point corresponds to a pair $(\mu^{DFT}_{\alpha\alpha},\mu^{GW}_{\alpha\alpha})$ in (a) and $(m^{*DFT}_{\alpha\alpha},m^{*GW}_{\alpha\alpha}),\, \alpha=1,\cdots 3$ in (b), as calculated for the strain configurations considered in Fig.~\ref{fig:best_case}. The brown lines indicate linear regression with zero $y$-intercept. All data are reported in Tables~\ref{tableIII} and \ref{tableIV} in Appendix, respectively.}
\label{fig:gw_dft}
\end{figure}

\subsection{Multi-component strain}\label{ssec.multi}

The single-component strain configurations considered in Secs.~\ref{ssec.summary}-\ref{ssec.mass} are not usually achievable in experiments, and in most cases multiple strain components are simultaneously nonzero. To address such general multi-component strain configurations, we estimate the mobility from the single-component strain data by employing the correlation between mobility and effective mass shown in Fig.~5. In Sec.~IV.D, we showed that the Drude-like parametrization in Eq.~(10) carries reasonable predictive power without performing full-blown \textit{ai}BTE calculations. This parametrization is advantageous in that: (i) it captures qualitatively the nonlinear effects of strain resulting from the cross-terms that are not included in Eq.~(1); (ii) it is suitable for rapid screening of strain configurations prior to full $ai$BTE calculations. Since this parametrization is obtained from a dataset of single-component strains, below we corroborate its validity by comparing with $ai$BTE calculations in several multi-component strain configurations.

The relation between strain components is found by minimizing the elastic energy under a given loading condition. In the case of wurtzite GaN considered here, the matrix of elastic constants in Voigt notation has five independent nonzero elements:\cite{Nye:1985,Polian:1996}
  \begin{equation}\label{eq.C}
    \renewcommand*{\arraystretch}{1.2}
    C=\begin{pmatrix}
      C_{11}\!\! & C_{12}\!\! & C_{13}\!\! &  & &  \\
      C_{12}\!\! & C_{11}\!\! & C_{13}\!\! &  &  & \\
      C_{13}\!\!&  C_{13}\!\!& C_{33}\!\! &  &  &  \\
      &  &  & \!C_{44}\!\! &  &  \\
      &  &  &  & \!C_{44}\!\! &  \\
      &  &  &  &  & \!C_{66} 
    \end{pmatrix},
  \end{equation}
with $C_{66}=(C_{11}-C_{12})/2$. The internal elastic energy per unit volume $U$ can be expressed using this matrix and the shear strain as [see Eq.~(6.15) of Ref.~\citenum{Giustino_2014}]:
  \begin{eqnarray}
    U &=& C_{11}(\epsilon_1^2+\epsilon_2^2+2\epsilon_6^2)/2+C_{12}(\epsilon_1\epsilon_2-\epsilon_6^2)
         \nonumber \\
    &+&  C_{13} (\epsilon_1\epsilon_3+\epsilon_2\epsilon_3) + C_{33}\epsilon_3^2/2
        + 2C_{44}(\epsilon_4^2+\epsilon_5^2).\label{eq.elastic}
 \end{eqnarray}
From this expression we see that the axial strains $\e_1$, $\e_2$, and $\e_3$ are coupled to each other by the bilinear terms $\e_1\e_2$, $\e_1\e_3$, and $\e_2\e_3$, while the shear strains $\e_4$, $\e_5$, and $\e_6$ are each decoupled from all other strain components. Therefore, if we set a single-component axial strain and minimize the elastic energy, additional axial strain will be present; conversely, if we impose a single-component shear strain, the ground state structure will not acquire additional strain components. This observation implies that, on the one hand, all our mobility results relating to shear strain directly map into realistic experimental conditions; on the other hand, an adjustment is needed for axial strains.

For axial strains, there are two inequivalent scenarios to consider. First, we focus on compressive uniaxial strain $\e_3$ induced by an applied stress in the same direction. Minimization of Eq.~\eqref{eq.elastic} yields:
  \begin{equation}
  \e_1 = \e_2 = -\frac{C_{13}}{C_{11}+C_{12}}\e_{3} , \quad
   \e_4 = \e_5 = \e_6 = 0 .\label{eq.e3}
  \end{equation}
Using our calculated elastic constants reported in Table~\ref{tab:Cij}, we obtain $C_{13}/(C_{11}+C_{12})=0.2$, therefore a 2\% compressive strain along the $c$ axis generates a 0.4\% tensile biaxial strain in the $ab$ plane of GaN, as expected from the Poisson ratio.\cite{Moram:2007} 
The room-temperature hole mobility corresponding to this scenario amounts to
$\mu_{zz}=164$\,cm$^2$/Vs
from the \textit{ai}BTE calculation. 
Using the conductivity effective mass of $m^*_{zz}=0.39~m_e$, Eq.~(10) yields the mobility $\mu_{zz}=215$\,\,cm$^2$/Vs, which is only 30\% higher than the \textit{ai}BTE result.

\begin{table}
\caption{Elastic constants of wurtzite GaN: Comparison between this work and literature data.}
\label{tab:Cij}
\begin{tabular}{ p{3.5cm} m{0.7cm} m{0.7cm} m{0.7cm} m{0.7cm} m{0.7cm} m{0.7cm} }
\hline\hline\\[-6pt]
Elastic constants (GPa) & $C_{11}$ & $C_{12}$ & $C_{13}$ & $C_{33}$ & $C_{44}$ & $C_{66}$ \\[4pt]
This work & 368 & 138 & 101 & 408 & \phantom{0}97 & 145\\[3pt]
Expt., Ref.~\citenum{Yamaguchi_1997} & 365 & 135 & 114 & 381 & 109 & 115 \\
Expt., Ref.~\citenum{Polian:1996} & 390 & 145 & 106 & 398 & 105 & 123 \\[3pt]
\hline\hline
\end{tabular}
\end{table}

Second, we consider compressive uniaxial strain $\e_1$ induced by an applied stress in the same direction. This time the minimization of Eq.~\eqref{eq.elastic} yields:
  \begin{equation}
  \e_2 = -\frac{C_{33}C_{12}-C_{13}^2}{C_{11}C_{33}-C_{13}^2}\e_{1}, \quad
  \e_3 = -\frac{C_{13}(C_{11}-C_{12})}{C_{11}C_{33}-C_{13}^2}\e_{1},\label{eq.e1} 
  \end{equation}
while the shear components vanish as in Eq.~\eqref{eq.e3}. In this case, the elastic constants from Table~\ref{tab:Cij} yield $\e_2 =-$0.3$\e_1$ and $\e_3=-$0.2$\e_1$, therefore a uniaxial compressive strain of 2\% along $x$ is accompanied by 0.6\% and 0.4\% dilations in $y$ and $z$, respectively. 
The resulting room-temperature hole mobility from $ai$BTE calculation is $\mu_{xx}=133$\,cm$^2$/Vs; using the conductivity effective mass $m^*_{xx}=0.37~m_e$, Eq.~(10) yields $\mu_{xx}=226$\,cm$^2$/Vs, which is about 70\% higher than the $ai$BTE mobility.

To test the model for multi-component strain configurations of practical interest, we consider the two configurations investigated in Ref.~\citenum{Leveillee:2022}, which correspond to GaN epitaxially matched to ZnGeN$_2$ and MgSiN$_2$, respectively.
In the case of GaN matched to ZnGeN$_2$, the strain is $\e_1 =-0.21\%$, $\e_2 =0.57\%$, $\e_3 =-0.1\%$, $\e_6=0.68\%$, and the conductivity effective masses are $m^*_{xx}=1.81~m_e$, $m^*_{yy}=0.60~m_e$, and $m^*_{zz}=1.77~m_e$. Equation~(10) yields the mobilities $\mu_{xx}=55$\,cm$^2$/Vs, $\mu_{yy}=146$\,cm$^2$/Vs, $\mu_{zz}=56$\,cm$^2$/Vs; these values compare well with our \textit{ai}BTE calculations for the same strain configuration, $\mu_{xx}=39$\,cm$^2$/Vs, $\mu_{yy}=100$\,cm$^2$/Vs, $\mu_{zz}=45$\,cm$^2$/Vs. 
In the case of GaN matched to MgSiN$_2$, the strain is $\e_1 =-2.82\%$, $\e_2 =0.21\%$, $\e_3 =0.6\%$, $\e_6=2.62\%$, and the conductivity effective masses are $m^*_{xx}=1.62~m_e$, $m^*_{yy}=0.36~m_e$, and $m^*_{zz}=2.20~m_e$. Equation~(10) yields $\mu_{xx}=60$\,cm$^2$/Vs, $\mu_{yy}=230$\,cm$^2$/Vs, $\mu_{zz}=46$\,cm$^2$/Vs; these values compare fairly with our \textit{ai}BTE calculations, $\mu_{xx}=48$\,cm$^2$/Vs, $\mu_{yy}=131$\,cm$^2$/Vs, $\mu_{zz}=46$\,cm$^2$/Vs, except for $\mu_{yy}$ which is overestimated by 70\%. For completeness, we indicate these data points in Fig.~5 as yellow squares. 
While this effective mass model yields reasonable trends, we emphasize that, to achieve predictive power, ultimately one needs to perform full-blown \textit{ai}BTE calculations, and that the piezomobility tensor and the effective masses only serve as useful pre-screening descriptors.

\section{Discussion and conclusions}\label{sec.concl}

In this work, we introduced the piezomobility tensor as a simple and effective conceptual tool for strain-engineering carrier mobilities in semiconductors via systematic \textit{ab initio} transport calculations. In the case of GaN, this approach allowed us to confirm a previously identified $c$-axis uniaxial strain as a viable route to increase the hole mobility, and to identify two additional strain conditions that should lead to a significant enhancement of the hole mobility, namely uniaxial $x$-axis compression and $yz$ shear strain. In all cases, we find that single-component strain between 2\% and 3\% can lead to hole mobilities reaching between 150 and 200\,\mob\ at room temperature (Fig.~\ref{fig:best_case}). These high mobilities are also found when considering more practical multi-component strain configurations which take into account the elastic deformation of the crystal (Sec.~\ref{ssec.multi}).

In all cases, the increase in hole mobility can be ascribed directly to a decrease of the conductivity effective mass of holes (Fig.~\ref{fig:effm}). For example, in the case of $x$-axis strain, the heavy holes and the split-off holes are pushed well below the light holes, leading to a decrease of the effective mass. Conversely, in the case of shear strain, it is the light holes and the split-off holes that are pushed down, leaving the heavy holes at the valence band top; however, the heavy hole band is heavily warped and the net effect is a reduction of the effective mass.

The present analysis paints a more complex picture of strain effects on the hole mobility of GaN than previously thought.\cite{Ponce:2019_2, Ponce:2019} To simplify, we could say that strain affects the hole mobility of GaN in one of two ways: (i) by changing the order of the topmost three valence bands and (ii) by modifying the corresponding band effective masses. Depending on the strain configuration, one of these two effects may be more pronounced than the other. For example, in the case of 2\% $\epsilon_4$ shear strain the effective mass along the \textit{c}-axis is reduced, without reversing the sign of the crystal-field splitting; in this case, the origin of the mobility enhancement lies in the change of effective mass. Conversely, if we consider 2\% $\epsilon_3$ compressive strain, the mobility enhancement results from the sign reversal of the crystal-field splitting, as reported in Ref.~\citenum{Ponce:2019_2}. In both cases we have a reduction of band effective mass, but for different bands. 

Strain levels considered in this work are experimentally feasible. Recently, several groups have realized strained GaN in quantum well heterostructures by epitaxial matching.  For example, the authors of Ref.~\citenum{Gupta:2019} employed InGaN in a fin geometry to achieve a 25-50\% reduction in sheet resistance under uniaxial compression, as compared to biaxially-strained layers.  Furthermore, in Ref.~\citenum{Jena2019}, undoped GaN was epitaxially grown on AlN to form a polarization-induced 2D hole gas.  Using reciprocal-space X-ray maps, the authors demonstrated that the epitaxial matching induces a 2.4\% biaxial compressive strain in the GaN layer. While we do not expect mobility enhancement for biaxial compression (see Fig.~\ref{fig:best_case}), this example illustrates the experimental feasibility of these strain conditions. 
In addition to the above, recently Ref.~\citenum{Wang2024} reported the experimental realization of extreme uniaxial compressive strain, $\epsilon_3 = -10\%$,  in $p$-type GaN via intercalation of two-dimensional Mg layers via the annealing a Mg film on GaN. These extremely strained GaN superlattices are expected to exhibit enhanced carrier transport properties; however, further experimental validation and a detailed understanding of the local atomic environment in the proximity of Mg layers will be necessary to fully understand the impact of doping and strain in these superlattices.

The mobility values reported in this work correspond to the phonon-limited mobility, and scattering mechanisms other than phonons, such as impurity scattering and grain boundary scattering, are not taken into account. For this reason, our calculations provide an upper bound to the mobility achievable in experiments. Furthermore, experimental measurements necessarily involve multiple strain components since the system tends to minimize the elastic energy. To investigate multi-component strain, we showed that a simple approach is to employ the correlation between the effective masses and the mobility to make predictions without resorting to explicit \textit{ab initio} mobility calculations (see Sec.~IV.E). However, we emphasize that these estimates should only be considered as a guideline for rapid screening, and full-blown \textit{ab initio} mobility calculations should be performed for accurate and predictive results.

In the case of the less common shear strain, the realization of $\epsilon_4$ shear strain (i.e. $yz$ shear) should also be possible by combining compressive and tensile strains along the principal axes of the shear deformation.\cite{Liu:2022} In fact, upon diagonalization of the strain tensor, one sees that these principal axes lie in the $yz$ plane, at $45^\circ$ from both the $y$ and $z$ axes.

In addition to our fully \textit{ab initio} approach, the ${\bf k}\!\cdot\!{\bf p}$ method is also frequently adopted to study the strain effect on the electronic structures of materials. In comparison with DFT or GW quasiparticle calculations, several works have reported faithful band structures with the parametrization of the low-dimension ${\bf k}\!\cdot\!{\bf p}$ Hamiltonian.}\cite{chuang1996,Kim1997,dugdale2000,Rinke2008} In Ref.~\citenum{Bader:2019}, the authors investigate the band structures of a few strain configurations using the ${\bf k}\!\cdot\!{\bf p}$ method and show that a uniaxial compression optimizes hole mobility in the GaN/AlN 2D hole gas. One limitation of the ${\bf k}\!\cdot\!{\bf p}$ method is that one needs to make assumptions about how phonons and electron-phonon couplings are modified by strain. For quantitative predictions of these effects, state-of-the-art first-principles mobility calculations, as presented in this work, need to be employed.

In summary, this work provides a systematic and comprehensive approach to designing high-hole-mobility GaN via strain engineering. More generally, the methodology presented in this work offers a general framework for first-principles-based mobility engineering through strain manipulation.

\acknowledgments\vspace{-10pt}
We gratefully acknowledge fruitful discussions with Grace Xing and Debdeep Jena. This research is supported by SUPREME, one of seven centers in JUMP 2.0, a Semiconductor Research Corporation (SRC) program sponsored by DARPA (Mobility calculations); and by the Computational Materials Science program of U.S. Department of Energy, Office of Science, Basic Energy Sciences under Award DE-SC0020129 (\texttt{EPW} software development). Computational resources were provided by the National Energy Research Scientific Computing Center (a DOE Office of Science User Facility supported under Contract No. DE-AC02-05CH11231), the Argonne Leadership Computing Facility (a DOE Office of Science User Facility supported under Contract DE-AC02-06CH11357), and the Texas Advanced Computing Center (TACC) at The University of Texas at Austin.

\section*{appendix}
In this Appendix we provide tables showing the hole mobility tensor and conductivity effective mass tensor, and a figure showing the phonon dispersions of strained GaN.
\begin{table*}
\caption{Hole mobility tensor of strained GaN, computed for a carrier concentration of 10$^{18}$\,cm$^{-3}$ and room temperature. Data outside of the brackets correspond to GW calculations, those inside the brackets are DFT-LDA calculations. The piezomobility tensor element $K_{ij}$ can be calculated by Eq. (1). For single component strain configurations, Eq. (1) gives the relation: $K_{ij}=\Delta\mu_i(\epsilon_j)/\epsilon_j.$}
\vspace{4pt}
\centering
\begin{longtable*}{@{\extracolsep{\fill}} r r r r r r r r r r r r r r r}
\hline\\[-8.5pt] \hline\\[-5pt]
\multicolumn{1}{l}{$\e$}&\multicolumn{1}{l}{$\e_i$}&\multicolumn{6}{c}{$\mu_{\alpha\alpha}$ (cm$^2$/Vs)}&\multicolumn{6}{c}{$\mu_{\alpha\beta}$ (cm$^2$/Vs)} \\[2pt]
&&$xx$&&$yy$&&$zz$&&$yz$&&$xz$&&$xy$&\\[2pt]
\cline{1-2} \cline{3-8} \cline{9-15}\\[-7pt]
0\%  &       &  57.8  & (45.0) & 57.8 & (45.0) &  47.7 & (36.8) &  &  &  &  &  & \\
-3\% & $\e_1$&  169.9 & (114.3)& 49.0 & (32.8) &  54.1 & (38.9) &  &  &  &  &  & \\
     & $\e_2$&  59.4  & (37.2) & 160.4& (108.5)&  53.2 & (38.4) &  &  &  &  &  & \\
     & $\e_3$&  67.1  & (50.3) & 67.1 & (50.3) &  214.0& (197.6)&  &  &  &  &  & \\
     & $\e_4$&  57.9  & (41.2) & 134.4& (93.2) &  188.0& (150.9)& 52.1 & (39.7) &  &  &  &\\
     & $\e_5$&  125.7 & (94.6) &  52.0& (40.7) &  182.7& (152.8)&  &  & 49.0 & (38.5) &  & \\
     & $\e_6$&  117.6 & (80.0) & 166.5& (104.3)&  52.3 & (37.7) &  &  &  &  & 81.6 & (54.1) \\
-2\% & $\e_1$&  158.6 &  & 48.2 &  & 52.7 &  &  &  &  &  &  & \\
     & $\e_2$&  55.5  &  & 151.2&  & 52.1 &  &  &  &  &  &  & \\
     & $\e_3$&  58.8  &  & 58.8 &  & 164.2&  &  &  &  &  &  & \\
     & $\e_4$&  57.4  &  & 119.6&  & 148.4&  & 35.4&  & &  &  & \\
     & $\e_5$&  111.9 &  & 52.1 &  & 143.5&  &  &  & 32.8&  &  & \\
     & $\e_6$&  112.4 &  & 147.2&  & 53.0 &  &  &  &  &  & 71.6& \\
-1\% & $\e_1$&  118.4 & (83.4) & 51.2 & (34.0) &  50.7 & (37.2) &  &  &  &  &  & \\
     & $\e_2$&  54.3  & (35.6) & 115.4& (77.8) &  50.6 & (36.4) &  &  &  &  &  & \\
     & $\e_3$&  52.5  & (41.5) & 52.5 & (41.5) &  63.1 & (58.5) &  &  &  &  &  & \\
     & $\e_4$&  59.8  & (43.5) & 81.2 & (60.0) &  81.7 & (62.9) & 7.9 & (8.8) &  &  &  &  \\
     & $\e_5$&  73.9  & (58.2) & 55.1 & (44.0) &  78.3 & (62.0) &  &  & 5.9 & (7.6) &  & \\
     & $\e_6$&  98.5  & (69.0) & 118.2& (79.2) & 51.8  & (37.5) &  &  &  &  & 54.6 & (39.5) \\
1\%  & $\e_1$&  50.4  & (34.1) & 101.0& (72.5) &  52.5 & (40.3) &  &  &  &  &  & \\
     & $\e_2$&  106.6 & (76.5) & 47.6 & (33.5) &  52.6 & (41.3) &  &  &  &  &  & \\
     & $\e_3$&  62.6  & (49.6) & 62.6 & (49.6) &  48.0 & (36.9) &  &  &  &  &  & \\
     & $\e_4$&  60.6  & (43.3) & 79.1 & (60.8) &  77.1 & (63.6) & -6.0 & (-6.5) &  &  &  & \\
     & $\e_5$&  72.9  & (58.0) & 55.7 & (43.5) &  75.9 & (61.9) &  &  & -5.1 & (-6.7) &  & \\
     & $\e_6$&  100.2 & (69.8) & 121.5& (80.0) &  52.5 & (38.0) &  &  &  &  & -55.1& (-39.7) \\
2\%  & $\e_1$&  52.9  &  & 122.8&  & 58.3 &  &  &  &  &  &  & \\
     & $\e_2$&  135.7 &  & 46.9 &  & 57.2 &  &  &  &  &  &  & \\
     & $\e_3$&  62.3  &  & 62.3 &  & 48.3 &  &  &  &  &  &  & \\
     & $\e_4$&  57.8  &  & 119.3&  & 144.5&  & -32.8&  &  &  &  & \\
     & $\e_5$&  112.4 &  & 52.4 &  & 141.5&  &  &  & -31.6&  &  & \\
     & $\e_6$&  114.5 &  & 150.6&  & 53.7 &  &  &  &  &  & -72.3&  & \\
3\%  & $\e_1$&  55.4  & (36.5) & 128.3& (93.6) &  59.7 & (48.1) &  &  &  &  &  & \\
     & $\e_2$&  155.7 & (108.6)&  47.5& (32.5) &  56.4 & (45.1) &  &  &  &  &  & \\
     & $\e_3$&  62.6  & (49.4) &  62.6& (49.4) &  47.5 & (36.8) &  &  &  &  &  & \\
     & $\e_4$&  58.4  & (40.8) & 134.4& (94.6) &  186.2& (153.5)& -49.2 & (-36.0) &  &  &  & \\
     & $\e_5$&  127.9 & (95.2) &  52.7& (40.6) &  182.4& (153.7)&  &  & -48.2 & (-37.6) &  & \\
     & $\e_6$&  119.3 & (78.7) & 168.3& (103.4)&  53.0 & (38.0) &  &  &  &  & -81.5& (-53.2) \\[5pt]
\hline\\[-8.5pt] \hline\\[-7pt]
\addtocounter{table}{-1} 
\end{longtable*}
\label{tableIII}
\end{table*}

\begin{table*}
\caption{Conductivity effective mass tensor of strained GaN, computed for a carrier concentration of 10$^{18}$\,cm$^{-3}$ and room temperature. Data outside of the brackets correspond to GW calculations, those inside the brackets are DFT-LDA calculations.}
\vspace{4pt}
\centering
\begin{longtable*}{@{\extracolsep{\fill}} r r r r r r r r r r r r r r r}
\hline\\[-8.5pt] \hline\\[-5pt]
&&$xx$&&$yy$&&$zz$&&$yz$&&$xz$&&$xy$&\\[2pt]
\cline{1-2} \cline{3-8} \cline{9-15}\\[-7pt]
0\%  &        & 1.32 & (1.39) & 1.32 & (1.39) & 1.65 & (1.90) &  &  &  &  &  & \\
-3\% & $\e_1$ & 0.34 & (0.34) & 2.12 & (2.40) & 2.07 & (2.32) &  &  &  &  &  & \\
     & $\e_2$ & 1.61 & (1.62) & 0.41 & (0.32) & 2.08 & (2.19) &  &  &  &  &  & \\
     & $\e_3$ & 1.65 & (1.84) & 1.65 & (1.84) & 0.38 & (0.37) &  &  &  &  &  & \\
     & $\e_4$ & 1.96 & (2.20) & 0.84 & (0.95) & 0.63 & (0.70) &-0.35 & (-0.42) &  &  &  & \\
     & $\e_5$ & 0.88 & (0.99) & 2.14 & (2.40) & 0.64 & (0.69) &  &  &-0.35 & (-0.42) &  & \\
     & $\e_6$ & 1.35 & (1.45) & 0.90 & (1.05) & 2.19 & (2.32) &  &  &  &  &-0.78 & (-0.94) \\
-2\% & $\e_1$ & 0.41 &  & 2.00 &  & 2.01 &  &  &  &  &  &  & \\
     & $\e_2$ & 1.62 &  & 0.47 &  & 2.02 &  &  &  &  &  &  & \\
\textbf{}     & $\e_3$ & 1.64 &  & 1.64 &  & 0.52 &  &  &  &  &  &  & \\
     & $\e_4$ & 1.72 &  & 0.86 &  & 0.75 &  &-0.31 &  &  &  &  & \\
     & $\e_5$ & 0.91 &  & 1.85 &  & 0.76 &  &  &  &-0.31 &  &  & \\
     & $\e_6$ & 1.31 &  & 0.93 &  & 2.08 &  &  &  &  &  &-0.75 &  & \\
-1\% & $\e_1$ & 0.59 & (0.73) & 1.58 & (2.03) & 1.90 & (2.09) &  &  &  &  &  & \\
     & $\e_2$ & 1.46 & (1.77) & 0.63 & (0.74) & 1.89 & (2.06) &  &  &  &  &  & \\
     & $\e_3$ & 1.50 & (1.59) & 1.50 & (1.59) & 1.16 & (1.25) &  &  &  &  &  & \\
     & $\e_4$ & 1.32 & (1.50) & 1.03 & (1.19) & 1.14 & (1.25) &-0.17 & (-0.11) &  &  &  & \\
     & $\e_5$ & 1.11 & (1.24) & 1.39 & (1.60) & 1.15 & (1.28) &  &  &-0.15 & (-0.15) &  & \\
     & $\e_6$ & 1.25 & (1.38) & 1.01 & (1.21) & 1.95 & (2.15) &  &  &  &  &-0.66 & (-0.80) \\
1\%  & $\e_1$ & 1.62 & (1.86) & 0.77 & (0.88) & 1.61 & (1.78) &  &  &  &  &  & \\
     & $\e_2$ & 0.71 & (0.85) & 1.74 & (2.07) & 1.62 & (1.79) &  &  &  &  &  & \\
     & $\e_3$ & 1.19 & (1.23) & 1.19 & (1.23) & 1.84 & (2.05) &  &  &  &  &  & \\
     & $\e_4$ & 1.30 & (1.50) & 1.06 & (1.19) & 1.20 & (1.25) & 0.16 & (0.11) &  &  &  & \\
     & $\e_5$ & 1.13 & (1.24) & 1.38 & (1.60) & 1.18 & (1.28) &  &  & 0.15 & (0.15) &  & \\
     & $\e_6$ & 1.23 & (1.38) & 0.98 & (1.21) & 1.96 & (2.15) &  &  &  &  & 0.64 & (0.80) \\
2\%  & $\e_1$ & 1.71 &  & 0.66 &  & 1.47 &  &  &  &  &  &  & \\
     & $\e_2$ & 0.56 &  & 1.97 &  & 1.52 &  &  &  &  &  &  & \\
     & $\e_3$ & 1.15 &  & 1.15 &  & 1.90 &  &  &  &  &  &  & \\
     & $\e_4$ & 1.69 &  & 0.87 &  & 0.77 &  & 0.31 &  &  &  &  & \\
     & $\e_5$ & 0.91 &  & 1.83 &  & 0.77 &  &  &  & 0.31 &  &  & \\
     & $\e_6$ & 1.27 &  & 0.91 &  & 2.09 &  &  &  &  &  & 0.72 &  & \\
3\%  & $\e_1$ & 1.69 & (1.75) & 0.64 & (0.83) & 1.39 & (1.55) &  &   &   &  &   &  \\
     & $\e_2$ & 0.49 & (0.61) & 2.05 & (2.34) & 1.52 & (1.78) &  &    &  &   &   & \\
     & $\e_3$ & 1.12 & (1.12) & 1.12 & (1.12) & 1.93 & (2.07) &  &    &  &   &  &  \\
     & $\e_4$ & 1.93 & (2.20) & 0.85 & (0.95) & 0.64 & (0.70) & 0.35 & (0.42) &  &  &  & \\
     & $\e_5$ & 0.87 & (0.99) & 2.12 & (2.40) & 0.64 & (0.69) &  &  & 0.35 & (0.42) &  & \\
     & $\e_6$ & 1.32 & (1.45) & 0.88 & (1.05) & 2.19 & (2.32) &  &  &  &  & 0.76 & (0.94) \\[5pt]
\hline\\[-8.5pt] \hline\\[-7pt]
\addtocounter{table}{-1} 
\end{longtable*}
\label{tableIV} 
\end{table*}

\begin{figure*}
\includegraphics[width=0.8\textwidth]{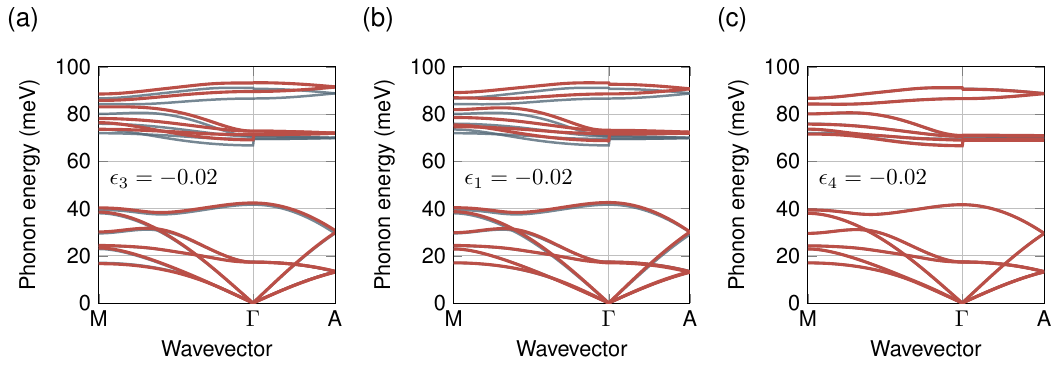}
\caption{Phonon dispersion relations of unstrained GaN (Gray) and strained GaN, for the three best case scenarios discussed in Sec.~IVB of the main text (brown): (a) 2\% compressive uniaxial single-component strain along the $c$ axis ($\e_3 = -0.02$); (b) 2\% compressive uniaxial single-component strain along $x$ axis ($\e_1 = -0.02$); (c) 2\% single-component $yz$ shear strain ($\e_4 = -0.02$).}
\label{fig:phSI}
\end{figure*}

\bibliography{biblio}

\end{document}